\RequirePackage{fix-cm}
\documentclass[smallextended,natbib]{svjour3} 

\smartqed
\usepackage[]{natbib}
\usepackage{xcolor}
\usepackage{graphicx}
\usepackage{amssymb}
\usepackage{lineno}
\usepackage{url}
\usepackage{bm}
\usepackage{framed}
\journalname{Neuroinformatics}

\usepackage{listings}
\definecolor{mygreen}{RGB}{28,172,0} 
\definecolor{mylilas}{RGB}{170,55,241}
\begin{document}
\title{Zeffiro user interface for electromagnetic brain imaging: a GPU accelerated FEM tool for forward and inverse computations in Matlab}
\titlerunning{Zeffiro user interface for electromagnetic brain imaging}
\author{Q.\ He
\and A.\ Rezaei
\and S.\ Pursiainen}
\institute{ Qin He \at 
             Information Technology, Faculty of Information Technology and Communication Sciences, Tampere University, P.O.\ Box 692, 33014 Tampere, Finland\\ 
\and Atena Rezaei (corresponding author)\at
             Mathematics and Statistics, Faculty of Information Technology and Communication Sciences, Tampere University, P.O.\ Box 692, 33014 Tampere, Finland, 
              \email{atena.rezaei@tuni.fi} \\
             \and Sampsa Pursiainen \at 
             Mathematics and Statistics, Faculty of Information Technology and Communication Sciences, Tampere University, P.O.\ Box 692, 33014 Tampere, Finland, 
              \email{sampsa.pursiainen@tuni.fi}\\ } 
\maketitle

\lstset{language=Matlab,%
    breaklines=true,%
    morekeywords={matlab2tikz},
    keywordstyle=\color{blue},%
    morekeywords=[2]{1}, keywordstyle=[2]{\color{black}},
    identifierstyle=\color{black},%
    stringstyle=\color{mylilas},
    commentstyle=\color{mygreen},%
    showstringspaces=false,
    emph=[1]{for,end,break},emphstyle=[1]\color{red}, 
}

\begin{abstract}

This article introduces the {\em Zeffiro}  interface (ZI) version 2.2 for brain imaging. ZI aims to provide a simple, accessible and multimodal open source platform for finite element method (FEM) based and graphics processing unit (GPU) accelerated forward and inverse computations in the Matlab environment. It allows one to (1) generate a given multi-compartment head model, (2) to  evaluate a lead field matrix as well as (3) to invert and analyze a given set of  measurements. GPU acceleration  is applied in each of the processing stages (1)--(3). In its current configuration, ZI includes forward solvers for  electro-/magnetoencephalography (EEG) and linearized electrical impedance tomography (EIT)  as well as a set of inverse solvers based on the hierarchical Bayesian model (HBM). We report the results of EEG and EIT inversion tests performed with real and synthetic data, respectively, and demonstrate numerically how the inversion parameters affect the EEG inversion outcome in HBM.  
The GPU acceleration was found to be essential in the generation of the FE mesh and the LF matrix in order to achieve a reasonable computing time. The code package can be extended in the future based on the directions given in this article. 
\end{abstract}



\maketitle

\section{Introduction}

This article introduces the {\em Zeffiro\footnote{{\em Zeffiro} is Italian for a {\em gentle breeze} referring to the ease of use. The source code of ZI can be accessed at: \url{https://github.com/sampsapursiainen/zeffiro_interface}.}} interface  (ZI) version 2.2 for electromagnetic brain imaging and investigations. ZI aims to provide an accessible and multi-modal open-source platform for finite element method (FEM)  \citep{braess2001} based forward and inverse computations in the Matlab (TheMathWorks Inc.) environment. The FEM is widely applied for modeling electromagnetic fields in a bounded domain, such as the brain and the head \citep{demunck2012,monk2003}. It allows one to discretize realistic three-dimensional tissue parameter distributions in an accurate way, including advanced features such as complex internal boundary layers and anisotropic tissues such as the fibrous white matter  of the brain \citep{rullmann2009eeg}. The FEM can be applied to model an electromagnetic source within the brain \citep{pursiainen2016,miinalainen2019} and, thereby, to construct a lead field (LF) matrix to localize brain activity in  electro-/magnetoencephalography (EEG/MEG)  \citep{hamalainen1993,niedermeyer2004}. 

The same quasi-static set of Maxwell's equations that predicts the electric potential field of a neural source can be applied also to model the effect of current injections, where either direct or alternating currents applied through electrodes act as the source of the electromagnetic field. Such an approach is used, for example, in the electrical impedance tomography (EIT) \citep{cheney1999electrical} in which the goal is to map the conductivity distribution or its perturbations within a given domain.  EIT constitutes a non-linear inverse problem which can be linearized  with respect to a given background conductivity  distribution to obtain a LF matrix, i.e., a linearized forward model. The FEM is a powerful tool in EIT \citep{vauhkonen1997}, since it does not set any major restrictions for the conductivity distribution. In contrast, the boundary element method (BEM) \citep{he1987electric}, which is the predominating method in EEG/MEG, sets the conductivity to be a compartment-wise constant parameter, limiting its practical usage in EIT. 

Until recently, the FEM has been considered as computationally heavy for discretizing the complex geometry of the brain. To tackle this issue, ZI uses graphics processing unit (GPU) acceleration. It includes forward solvers for EEG/MEG and linearized EIT as well as a set of inverse solvers based on  the hierarchical Bayesian model (HBM) which was introduced for EEG/MEG in \cite{calvetti2009}. The ZI platform and function library has been designed to be easily expandable and to allow implementing virtually any FEM based forward model which can be formulated as a product between a LF matrix and a candidate solution vector. 

In this paper, we briefly review the mathematics behind ZI, describe the  principal operations and usage, and introduce some central points for the developer perspective. We report the results obtained in EEG and EIT inversion tests performed with real and syntetic data, respectively, and demonstrate numerically how the inversion parameters affect the EEG inversion outcome in HBM. 




\section{Methodology}
\label{methods}

The electric potential field $u$ in the head model $\Omega$ is assumed to satisfy the elliptic partial differential equation (PDE) of the form $\nabla \cdot (\sigma \nabla u) = \nabla \cdot \vec{J}^{\, p}$, where $\sigma$ is the conductivity distribution of the head and $\vec{J}^{\, p}$ is the primary current density of the neural activity. This equation follows from the current preservation condition $\nabla \cdot \vec{J}^{\, t} = 0$ for the total current density  $\vec{J}^{\, t} =  \vec{J}^{\, p} - \sigma \nabla u$, that is, the sum of $\vec{J}^{\, p}$ and the volume current density $-\sigma \nabla u$. The electromagnetic field within $\Omega$ can be evoked either by $\vec{J}^{\, p}$ acting as the source, which is the case in EEG/MEG, or by an external source, e.g., a current pattern injected through contact electrodes in EIT. The dependence between the measurements ${\bf y}$ and the unknown of the inverse  problem ${\bf x}$ in question, e.g., a source localization problem, is here assumed be of the following linear form 
\begin{equation}
\label{linear_forward_model}
    {\bf L} {\bf x} = {\bf y} + {\bf n}, 
\end{equation} where ${\bf L}$ is the LF matrix and ${\bf n}$ is the noise vector. The LF matrices for EEG and linearized EIT inverse problem can be formed as shown in Appendix \ref{appendix_cem}.

\subsubsection{Primary current model}

 ZI utilizes the H(div) source model \citep{pursiainen2016} in which both linear and quadratic basis functions constitute the primary current density $\vec{J}^{\, p}$. In \cite{miinalainen2019,pursiainen2016}, this model was shown to surpass the accuracy of the classical direct source modeling approaches based on the partial integration and St.\ Venant's principle and to be especially advantageous for thin cortices as well as for inverting data.
 
A Cartesian set of source orientations can be obtained from a  mesh-based set using the Position Based Optimization (PBO) method \citep{bauer2015} with an adaptive \citep{miinalainen2017} 10-source stencil in which 4 face and 6 edge functions are applied for each element containing a source  \citep{pursiainen2016}. Alternatively, the Whitney model  \citep{bauer2015}, i.e., the 4-source stencil (4 face functions), can be used. Moreover, a set of Whitney functions can be applied  without interpolation. That is, the LF matrix can be formed directly using the mesh-based set of basis functions as suggested in \cite{miinalainen2017}. In each active tissue compartment, the sources can either be normally constrained or unconstrained  with respect to the  surface of the compartment  \citep{creutzfeldt1962influence,hari2018ifcn}.  The source positions are  randomly (uniformly) distributed in each case. 

\subsubsection{Conductivity distribution}

The current FE meshing strategy employed in ZI treats the conductivity as an isotropic piecewise (element-wise) constant distribution, i.e., a single scalar value is associated with each element in the FE mesh. However, when evaluating an LF matrix, ZI allows the conductivity distribution $\sigma$ to be anisotropic, i.e., tensor-valued:  the $\ell$-th row of the form $(\sigma_{11}, \sigma_{22}, \sigma_{33}, \sigma_{12}, \sigma_{13}, \sigma_{23})$ within a multi-row array is associated with the symmetric conductivity tensor $\sigma_{ij}$, $i=1,2,3$, $j =1,2,3$  ($\sigma_{ij} = \sigma_{ji}$)  in the $\ell$-th element. 

\subsection{HBM}

The  inverse tools of  ZI are based on  the HBM \citep{calvetti2009,ohagan2004} which enables finding a  reconstrution for the unknown ${\bf x}$ as either the posterior maximizer, i.e., {\em maximum a posteriori} (MAP) or the conditional mean (CM) of the {\em posterior} probability density.
In HBM, the posterior probability for ${\bf x}$ is defined via choosing the standard deviation of a Gaussian likelihood density, the hypermodel, i.e., the gamma (G) or inverse gamma (IG) hyperprior determining the actual prior, and the shape and scale parameter $\beta$ and $\theta_0$ for the hyperprior. For a given measurement vector ${\bf y}$, the Bayes formula \citep{ohagan2004} for the posterior is of the form
\begin{equation}
p ( {\bf x} \mid {\bf y})  =   \frac{p({\bf x}) \, p({\bf y} \mid {\bf x})} {p({\bf y})} \propto  {p({\bf x}) \, p({\bf y} \mid {\bf x})},
\label{posterior}
\end{equation} 
where  $p({\bf x})$ is the prior density and $p({\bf y} \mid {\bf x})$ the likelihood function  \citep{schmidt1999bayesian}. Here, the noise term ${\bf n}$, which together with the forward model (\ref{linear_forward_model}) implies the likelihood  $p({\bf y} \mid {\bf x})$, is assumed to be a Gaussian zero-mean random vector with independent entries. 

In HBM, the prior can be expressed in the following hierarchical form 
$
p ( {\bf x}, {\bf h}) \propto  p({\bm \theta}) \,  p ( {\bf x} \mid {\bm \theta}),
$ 
where ${\bm \theta}$ is the primary hyperparameter of the model. The conditional part $p ( {\bf x} \mid {\bm \theta})$ of the prior is  a zero-mean Gaussian density, whose diagonal covariance matrix is predicted by the hyperprior $p ({\bm \theta})$. The hyperprior is assumed to have a long-tailed density, implying that ${\bf x}$ is likely to be a sparse vector corresponding to a well-localized (focal) volumetric distribution. In ZI, it is either  G or  IG density \citep{calvetti2009}, which are controlled by the  shape and scale parameter $\beta$ and $\theta_0$. The G and IG hyperprior can be coupled into a single model in a straightforward way, since the reciprocal $\theta^{-1}$ of a G-distributed random variable $\theta$ with respect to $\beta$ and $\theta_0$ is IG-distributed w.r.t.\ $\beta$ and $\theta_0^{-1}$.

A description of the IAS algorithm applied in ZI can be found in Appendix  \ref{appendix_ias}. ZI's CM estimation technique is based on the Gibbs sampler algorithm  \citep{spitzer1971markov,murphy2012machine} according to \cite{calvetti2009}.

\subsection{Hardware requirements}

ZI is principally designed to be used with a workstation or a high-end desktop computer with tens of gigabytes of RAM,  a multi-core CPU and one or more GPUs. When generating the FE mesh and the LF matrix ZI is likely to allocate several gigabytes of RAM. A one-millimeter FE mesh resolution might lead to  64 GB of motherboard RAM and 2--4 GB of GPU RAM allocation during the forward computations. The resulting FE mesh will consist of 3-4 M nodes and 20-30 M elements, and the eventual project size, when stored on a hard disk, will be 0.5--1 GB.

\subsection{GPU function}


ZI utilizes a GPU to accelerate the FE mesh generation process, forward and inverse computations, source interpolation and decompositions, as well as to speed up 3D visualizations. 
This is vital in order to achieve a convenient, around one hour computing time for a one-millimeter FE mesh resolution which has been shown to be  essential in order to obtain  physiologically accurate inverse estimates \citep{rullmann2009eeg}. A GPU is a parallel processing unit which has somewhat limited  RAM compared to  the motherboard. It can handle computation intensive operations very effectively, while memory intensive operations should be avoided. The operations related to forward and inverse computations can be accelerated due to the fast processing of  matrix-vector products in a GPU. The other GPU operations are mainly based on the acceleration of {\tt find} and {\tt sort} routines, evaluating those as blocks rather than individual entries. 

\subsubsection{Forward simulation}

In the Matlab environment, the most essential speed-up gain is related to the sparse FE matrix-vector products which need to be evaluated iteratively in the forward simulation phase. The GPU-parallelization of the forward simulation is especially  important, because Matlab currently handles the sparse matrix products in a single processor thread.  To evaluate the lead field matrix as described in \ref{transfer_matrix}, ZI uses the preconditioned conjugate gradient (PCG) \citep{golub1989} method with a lumped diagonal preconditioner
(LDP) in which each diagonal entry is obtained as the row sum of the absolute entry values. LDP is an advantageous preconditioner regarding the limited GPU memory. While LDP  is not optimal with respect to minimizing the iteration steps needed for convergence, it enables establishing a fast forward solver due to the high parallel processing performance provided by a GPU. 

\subsubsection{IAS iteration}

In the IAS iteration (Appendix \ref{appendix_ias}), the most time consuming step is the third one, Equation  (\ref{ias_matrix}),  in which the size of the matrix to be inverted is determined by the length of the data vector. If a high number of time steps will need to be processed, the fastest processing is obtained by evaluating the matrix-vector product of (\ref{ias_matrix}) in a GPU. 

\section{Interface structure and function}

\begin{figure*}[h!]\begin{center}
\begin{scriptsize}
\includegraphics[width=10cm]{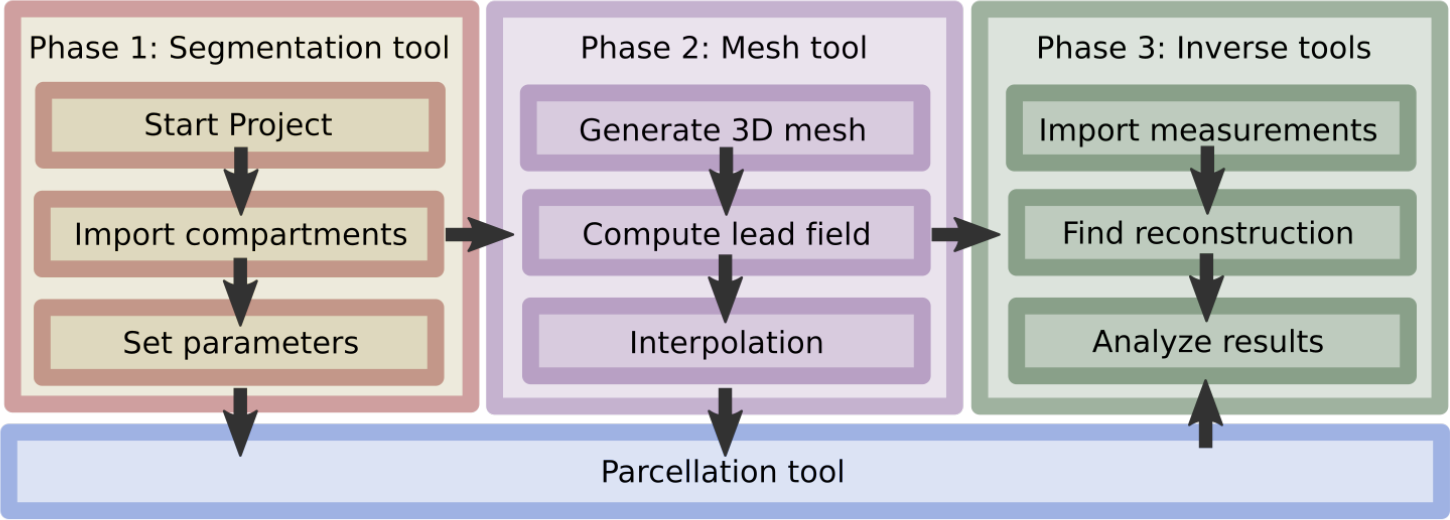}
\caption{The basic three-phase workflow in ZI. In phase 1, the head model is first defined using the {\em segmentation tool}, after which, in phase 2, the three-dimensional FE mesh and the LF matrix are generated with the {\em mesh tool}. Finally, in phase 3, the {\em inverse tools} can be applied to reconstruct and analyze parameter distributions, e.g., the primary current density of the brain activity. The {\em parcellation tool}  can be applied in each of the phases 1--3 to assist decomposing the brain into a finite set of ROIs. \label{workflow}}
\end{scriptsize}
\end{center}
\end{figure*}

When started, ZI creates a single data structure (struct) {\tt zef} in Matlab's base workspace. All the parameters and variables, such as the lead field matrix, measurement data and reconstruction, can be accessed via the {\tt zef} structure. The basic workflow consists of three phases  illustrated in Figure \ref{workflow}.  In this section, we briefly review the workflow and introduce the most important fields of {\tt zef} for each phase.


\begin{figure*}[h!]\begin{center}
\begin{scriptsize}
\includegraphics[width = 12cm]{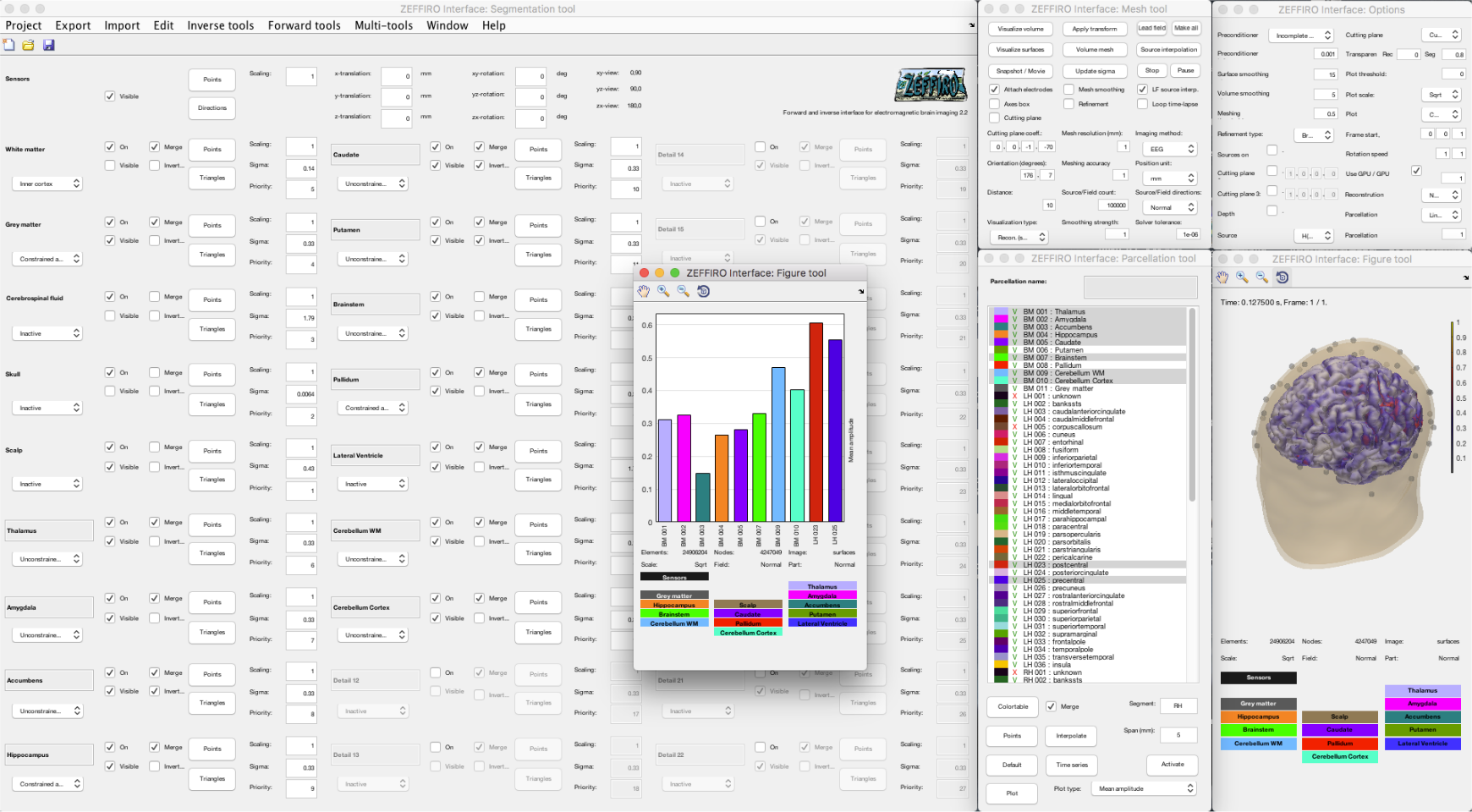} \\ 
\end{scriptsize}
\caption{A screenshot of ZI with figure, mesh, parcellation, and option tool opened. \label{main_window}}
\end{center}
\end{figure*}

\subsubsection{Segmentation tool}
\label{segmentation_tool}

In the first phase, a surface segmentation describing different  tissue structures and properties within $\Omega$ is defined using the {\em segmentation tool} (Figure \ref{main_window}). A triangular surface mesh for each tissue type is imported in ZI as an ASCII file. In the current version, a single head model can contain up to 27 different tissue compartments. Moreover, several surface meshes (sub-meshes) can be merged together into a single compartment, e.g., the left and right hemisphere of the cerebral cortex. A multi-compartment segmentation can be defined in a single initialization (.INI) file which allows importing a complete head segmentation at once. The nodes and points of each surface mesh can be stored either in two  separate .DAT files or in a single .ASC file exported from the FreeSurfer\footnote{\url{https://surfer.nmr.mgh.harvard.ed}} Software Suite \citep{fischl2012freesurfer}.

The default set of compartments includes white matter, grey matter, cerebrospinal fluid (CSF), skull, and scalp, whose default conductivity values are 0.14, 0.33, 0.0064, and 0.43 S/m, respectively, according to  \cite{dannhauer2010,vorwerk2014guideline}. Each compartment can be defined as active or inactive. The set of  active compartments contains the DOFs of ${\bf x}$. In EEG/MEG, the activity can be either constrained or unconstrained. In the former case, it is restricted into the direction of the surface normal, and in the latter case, it can have any orientation.


\subsubsection{Mesh tool}

\begin{figure}[h!]\begin{center}
\includegraphics[width = 8cm]{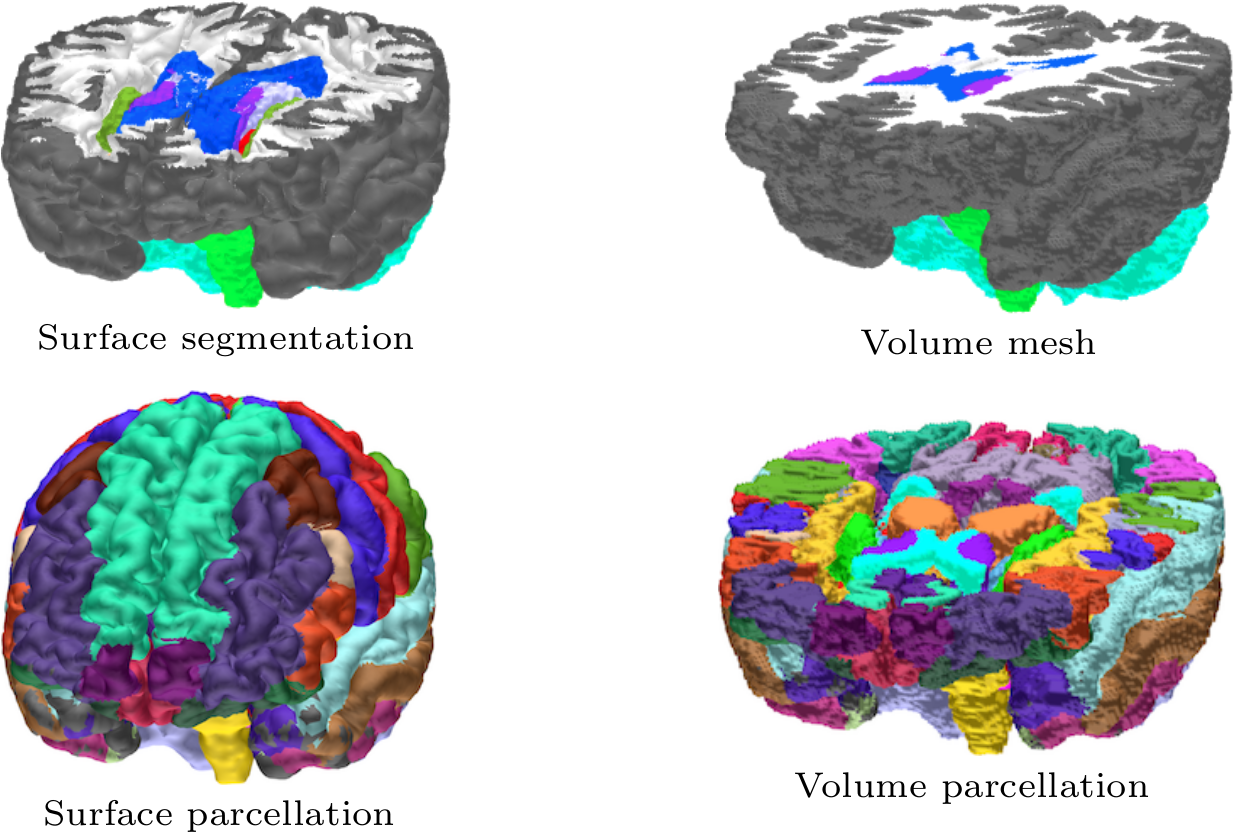}
\caption{{\bf Top row:} Surface and volume visualizations of the head model in ZI. {\bf Bottom row:} FreeSurfer-based cortical parcellation with 36 Desikan-Killiany labels. \label{reconstruction}}
\end{center}
\end{figure}

In the second phase of the workflow, a uniform tetrahedral mesh is generated based on the surface segmentation. The meshing parameters can be defined in the {\em mesh tool}. The meshing process proceeds from the innermost (detail) compartment to the outermost one. It allows the tissue boundaries to intersect each other which is necessary with a real segmentation obtained from magnetic resonance imaging (MRI) data. Each compartment can be given a priority which is referred to if a tetrahedron has nodes in two or more compartments. The priority parameter allows fine-tuning the width of the thin tissue layers, e.g., the skull: the lower the value the higher the priority. The FE mesh can be also smoothed using the Bi-Laplacian smoothing flow  \citep{ohtake2001,pursiainen2012raviart}. After generating the mesh, the LF matrix can be computed for a selected imaging modality and a given number of the degrees of freedom (DOFs). Finally, an interpolation process connecting the DOFs and the FE mesh nodes needs to be performed, to enable inversion of measurement data.

Together with the {\em figure tool},  the mesh tool allows one to visualize both the surface segmentation and the volumetric FE mesh, or any surface or volumetric distribution (reconstruction) defined on those. The visibility of a compartment can be selected  in the Segmentation tool. The {\em options tool} includes additional options which control, e.g., the colormap, scale, vector component, and the index of the sub-mesh  for the visualized distribution, e.g., that of the left or right hemisphere (see Section \ref{segmentation_tool}). An example of a multi-layer surface segmentation and the resulting volumetric mesh created with ZI are shown in Figure \ref{reconstruction}. For further code development, the most important fields of {\tt zef} are the following:
\begin{enumerate}
\item {\tt zef.nodes} and {\tt zef.tetra} store the nodes and tetrahedra of the FE mesh, respectively;
\item {\tt zef.L} is the lead field matrix; 
\item {\tt zef.source\_positions} stores the source positions corresponding to the columns of {\tt zef.L} in the respective order. This array contains the DOF positions also if they do not represent neural sources, which is the case in EIT. 
\item {\tt zef.source\_directions} contains the source orientations. If Cartesian orientations are used, this field is empty, and the source orientation for the columns of {\tt zef.L} is given by the following regular pattern: position 1, xyz; position 2, xyz; position 3, xyz, etc.;
\item {\tt zef.source\_interpolation\_ind} stores the indices that connect the finite element mesh with the DOFs;
\item  {\tt zef.h\_axes1} stores the axes handle of the figure tool.
\end{enumerate}
 
\subsubsection{Inverse tools}

In the third phase, the measurement data are imported and, after that, a reconstruction for ${\bf x}$ can be obtained using one of the {\em inverse tools}.  A MAP estimate can be obtained via the IAS method using one of the following tools: \begin{enumerate} \item  {\em IAS MAP estimation} which finds a MAP estimate for the whole domain; \item  {\em IAS MAP estimation ROI} which focuses on a ROI;  \item  {\em IAS MAP multiresolution} which explores multiple different resolutions. \end{enumerate} 
 A CM estimate can be obtained for a ROI using the {\em Hierarchical Bayesian sampler} tool. For external inverse procedure development, the most important fields are the following:
\begin{enumerate}
\item {\tt zef.measurements} is the set of measurements to be inverted;  this field can be a matrix or a cell array with the number of rows and columns equal to that of {\tt zef.L} and the time steps in the dataset, respectively;
\item {\tt zef.reconstruction} is the reconstruction of ${\bf x}$ corresponding to the set of source positions and  orientations.
\end{enumerate}

\subsubsection{Parcellation tool}

The {\em parcellation tool} (Figure \ref{main_window}) allows importing a parcellation created with the FreeSurfer Software Suite. A single parcellation consists of a file containing a colortable (.MAT) and another one including the points/labels  (.ASC).  After importing, an interpolation process will need to be performed to connect the points with the DOFs. The parcellation can be used as {\em a priori} information in the reconstruction or visualization stage. After obtaining a reconstruction, one can evaluate a time series of the activity for each region present in the parcellation. The time series can represent, e.g., the maximal or median activity within a region. The purpose of the  time series is to enable the analysis of different statistical  properties and connectivity of the activity over a time interval. In the current version, e.g., the amplitude, standard deviation, correlation, covariance, and dynamic time warping (DTW) \citep{sakoe1978} measure can be evaluated. The most important fields w.r.t.\ the parcellation tool are the following:
\begin{enumerate}
\item {\tt zef.parcellation\_colortable} and {\tt zef.parcellation\_points} store the colortable and points of the parcellation;
\item {\tt zef.parcellation\_interp\_ind} contains the indices connecting the parccellated brain regions and the DOFs;
\item {\tt zef.parcellation\_time\_series} stores the time series obtained for the brain regions after reconstructing the brain activity.
\end{enumerate}

\subsection{Plugin utility}

 ZI can be extended via the {\em plugin utility}. The list of plugins is defined in the {\tt zeffiro\_plugins.ini} file which is located in ZI's root folder. A menu item will be created for each listed plugin. The Hierarchical Bayesian sampler tool \citep{spitzer1971markov,murphy2012machine} is included in the code package as an example plugin (HBSampler). 

\subsection{Numerical experiments}

In the numerical experiments, we demonstrate the practical performance of ZI and the IAS MAP estimation technique via numerical experiments in which EEG and EIT inversion is tested  with real and synthetic data, respectively. We also analyze the effect of hyperprior and scale parameter on the source localization in EEG using simulated measurements.

\subsubsection{EEG inversion test}
\label{EEG_test}

\begin{figure}[h!]\begin{center}
\begin{scriptsize}
\begin{minipage}{9cm}
\begin{center}
\includegraphics[width = 9cm]{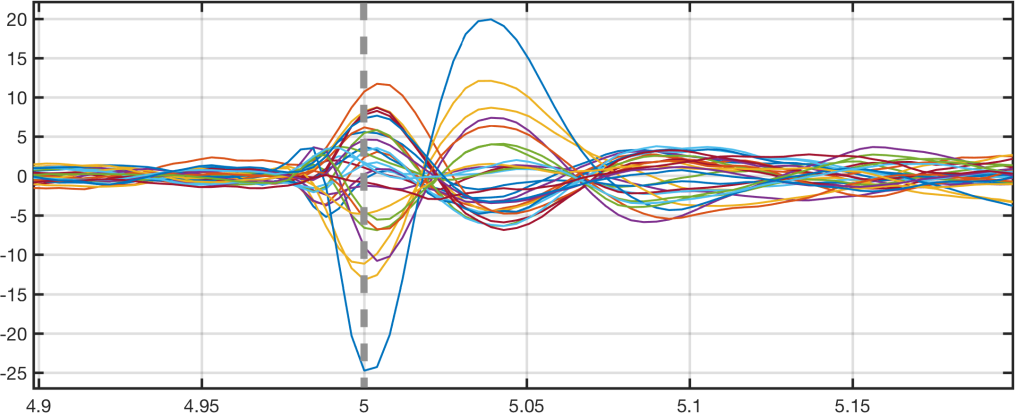}  \\
\end{center}
\end{minipage}  
\end{scriptsize}
\caption{A butterfly plot of EEG inversion test (Section \ref{EEG_test}) data which were obtained by averaging 58 epileptiform discharges  between -5 and 5 s. The vertical axis shows the measured voltage in microvolts, and the horizontal axis the measurement time in seconds. The reconstruction was found for the zero time point 0 s which is indicated  by the vertical dashed line. \label{butterfly_plot}}
\end{center}
\end{figure}

To enable comparability of the results to an existing solver, in this case that of the Brainstorm\footnote{\url{https://neuroimage.usc.edu/brainstorm/Introduction}}  software \citep{tadel2011brainstorm}, EEG source localization accuracy was examined by inverting Brainstorm's  {\em EEG and epilepsy} tutorial  dataset\footnote{\url{https://neuroimage.usc.edu/brainstorm/Tutorials/Epilepsy}} which was used with the consent of Prof.\ A.\ Schulze-Bonhage, Epilepsy Centre, University Hospital Freiburg, Germany. The dataset was obtained for a patient who had suffered from focal epilepsy with focal sensory, dyscognitive and secondarily generalized seizures since the age of eight years. It consists of 58 epileptiform discharges (spikes) which were recorded at 256 Hz frequency and detected using Brainstorm by the epileptologists in Freiburg. An invasive EEG study concentrating on the same subject can be found in \cite{dumpelmann2012sloreta}.

Following the tutorial, the measurement data to be inverted, depicted in Figure \ref{butterfly_plot}, were obtained for  29 electrodes applying an epoching time interval between -5 and 5 s w.r.t.\ the time point of the inverted data. All the non-EEG channels have been removed from the measured data. The brain activity was reconstructed via two steps of the IAS MAP estimation algorithm with low-cut and high-cut  frequency of 0.5 Hz and  80 Hz, assuming that the likelihood standard deviation is 3 \% of the maximum entry in the data, and selecting the shape and scale parameters as $\beta = 1.5$ and $\theta_0 =$ 1E-12. The head model linked to the dataset consists of the surface meshes of the scalp, skull, CSF, grey matter, and white matter. ZI's default conductivity values were used. The LF matrix was generated for 100000 sources using 1 mm mesh resolution. The reconstructions were obtained with ZI's {\em IAS MAP estimation} inverse tool.

\subsubsection{EIT inversion test}
\label{eit_test}

EIT inversion was examined numerically using the population head model\footnote{\url{https://itis.swiss/virtual-population/regional-human-models/phm-repository/}} which includes a scalp, skull, CSF, ventricle, grey matter, and white matter compartment  \citep{lee2016investigational}. The default conductivity values were applied, associating the condutivity of the ventricles with that of the CSF. A total of 72 ring electrodes with an assumed 1 kOhm impedance and an outer and inner diameter of 10 and 7.5 mm, respectively, were  modeled through the complete electrode model (CEM) described in Appendix \ref{appendix_cem}. 

\begin{figure}[h!]\begin{center}
\includegraphics[width = 10cm]{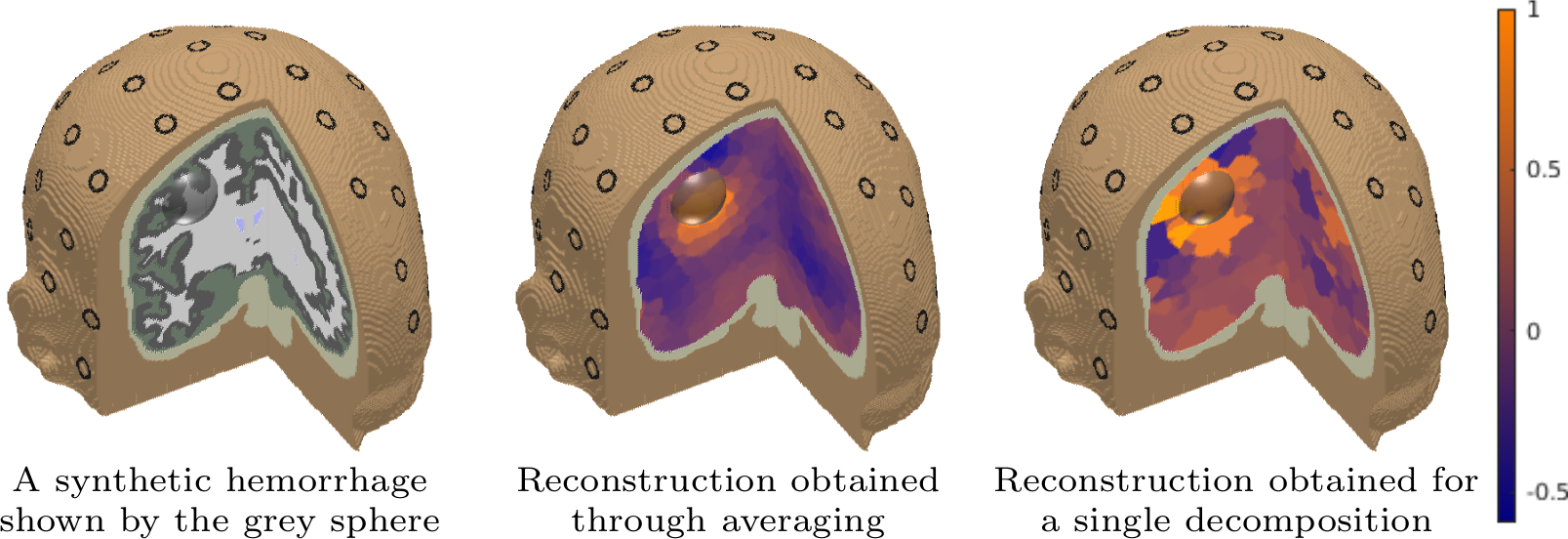}
\caption{{\bf Left:} An illustration of the synthetic hemorrhage (grey sphere) which was applied to generate the data of the EIT inversion tests. The diameter of the sphere was 30 mm and its conductivity was set to be 0.73 S/m higher compared to its surroundings. The unperturbed background conductivity distribution was assumed to be constant in each tissue compartment including white matter (white), grey matter (grey), CSF (green and blue), skull (khaki), and scalp (brown). The CEM electrodes (Appendix \ref{appendix_cem}) are shown as surface patches (black rings): {\bf Center:} An averaged  reconstruction of the synthetic hemorrhage found using the {\em IAS MAP multiresolution} inverse tool. The final distribution was produced as an average of altogether 100 different MAP estimates corresponding to different randomized decompositions of 300 DOFs as explained in Section \ref{eit_test}. {\bf Right:} A reconstruction (an unaveraged  MAP estimate) found for a single decomposition of 300 DOFs. The reconsructions have been normalized to one.  \label{eit_reconstructions}}
\end{center}
\end{figure}

The head model was discretized using  1 mm mesh resolution. The FE mesh is shown in Figure \ref{eit_reconstructions}.  A LF matrix was evaluated for a total number of 5000 DOFs using the approach presented in Appendix \ref{appendix_cem} and the original piecewise constant conductivity as the background distribution, i.e., the point of the linearization. The DOFs were distributed in the CSF, white matter and grey matter compartment. 

The synthetic data were generated by perturbing the conductivity inside the brain within a spherical 30 mm diameter sub-domain representing an intracerebral hemorrhage \citep{broderick1993volume}. Following, e.g., \cite{li2017construction,tang2010robust}, the magnitude of the perturbation was set to be +0.73 S/m and the signal-to-noise ratio was assumed to be 60 dB. The measurement errors consisted of additive Gaussian zero-mean white noise. 

The likelihood standard deviation was set to be 12 \% conciding approximately with the level following from the noise model. The IG hyperprior was employed selecting the shape and scale parameters as $\beta = 1.5$ and $\theta_0 =0.001$. To reconstruct the deep-lying anomaly, the total set of DOFs was decomposed into randomized 300 subsets which were formed w.r.t.\ an equal number of randomly (uniformly) distributed center points via the nearest point interpolation technique. The MAP estimate was found by performing two steps of IAS iteration for altogether 100 such randomized decompositions. A serial approach was adopted: the estimate obtained for one decomposition was set as the initial guess for the next one. The final reconstruction was produced as the mean of the resulting 100 MAP estimates. 

The motivation to use averaging was to reduce the effect of decomposition-related artifacts  which we assumed to be identically distributed for each separate decomposition and, thus, converge towards an expectation of an asymptotical Gaussian distribution based on  the law of random numbers and the central limit theorem \citep{ohagan2004}. The averaged reconstruction  was obtained using the {\em IAS MAP multiresolution} inverse tool which allows averaging the reconstruction over one or more resolution levels and multiple randomized decompositions. The resolution is determined by the number of subsets within a single decomposition which is here 300 in each.  

\subsection{Hypermodel and parameter selection}
\label{hypermodel_selection}

The HBM approach requires selecting the hypermodel together with an appropriate value for the shape and scale parameter $\beta$ and $\theta$. To investigate the effect of the parameter selection on the IAS MAP estimation process, we compared the localization of a simultaneously active pair of synthetic deep and superficial 10 nAm source in the case of EEG. The reconstruction was found as the center of mass of the primary current distribution within two 30 mm ROIs centered at the actual source locations. The accuracy was measured by evaluating the position (mm) and orientation (degree) difference with respect to the exact sources. As the computation domain we used a six-compartment (white matter, grey matter, CSF, compact skull, spongious skull, scalp) head model corresponding to a 49-year old male subject with ZI's default conductivity values. For the spongious part of the skull 0.028 S/m was selected \citep{vorwerk2014guideline}. The EEG LF matrix was formed for a cap of 72 electrodes. The effects of choosing the hyperprior $h$ and scale parameter $\theta_0$ were examined for the following four pars: {(\bf i)} $h=\hbox{G}, \theta_0 = \hbox{1E-5}$, {(\bf ii)} $h = \hbox{IG}, \theta_0 = \hbox{1E-5}$, {(\bf iii)} $h=\hbox{G}, \theta_0 = \hbox{1E-9}$, and {(\bf iv)} $h = \hbox{IG},\theta_0 = \hbox{1E-9}$, respectively.  The shape parameter $\beta$ was set to be $\beta = 1.5$ in each case. Gaussian white noise with 2 \% relative standard deviation was added in the data. Each reconstruction was evaluated for 50 different realizations of the noise vector. The inverse tool applied in the experiment was ZI's {\em IAS MAP estimation ROI}. 

\begin{figure}[h!]\begin{center}
\includegraphics[width = 9cm]{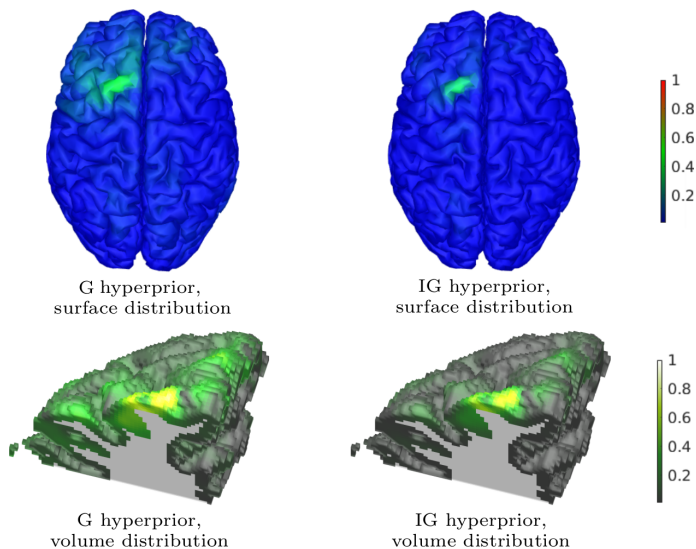}
\caption{A surface and volume visualization  of reconstructed brain activity (amplitude) obtained in the EEG inversion test (Section \ref{EEG_test}). The left and right images correspond to G and IG  hyperprior, respectively. {\bf Top row:} An axial projection of the reconstructions interpolated on the surface of the grey matter compartment. {\bf Bottom row:} The volumetric reconstructions cut by a coronal plane at the location of the maximal activity. The reconstructions have been normalized to one.  \label{eeg_reconstructions}  }
\end{center}
\end{figure}

\begin{figure}[h!]
   \begin{center}
 \includegraphics[width = 10cm]{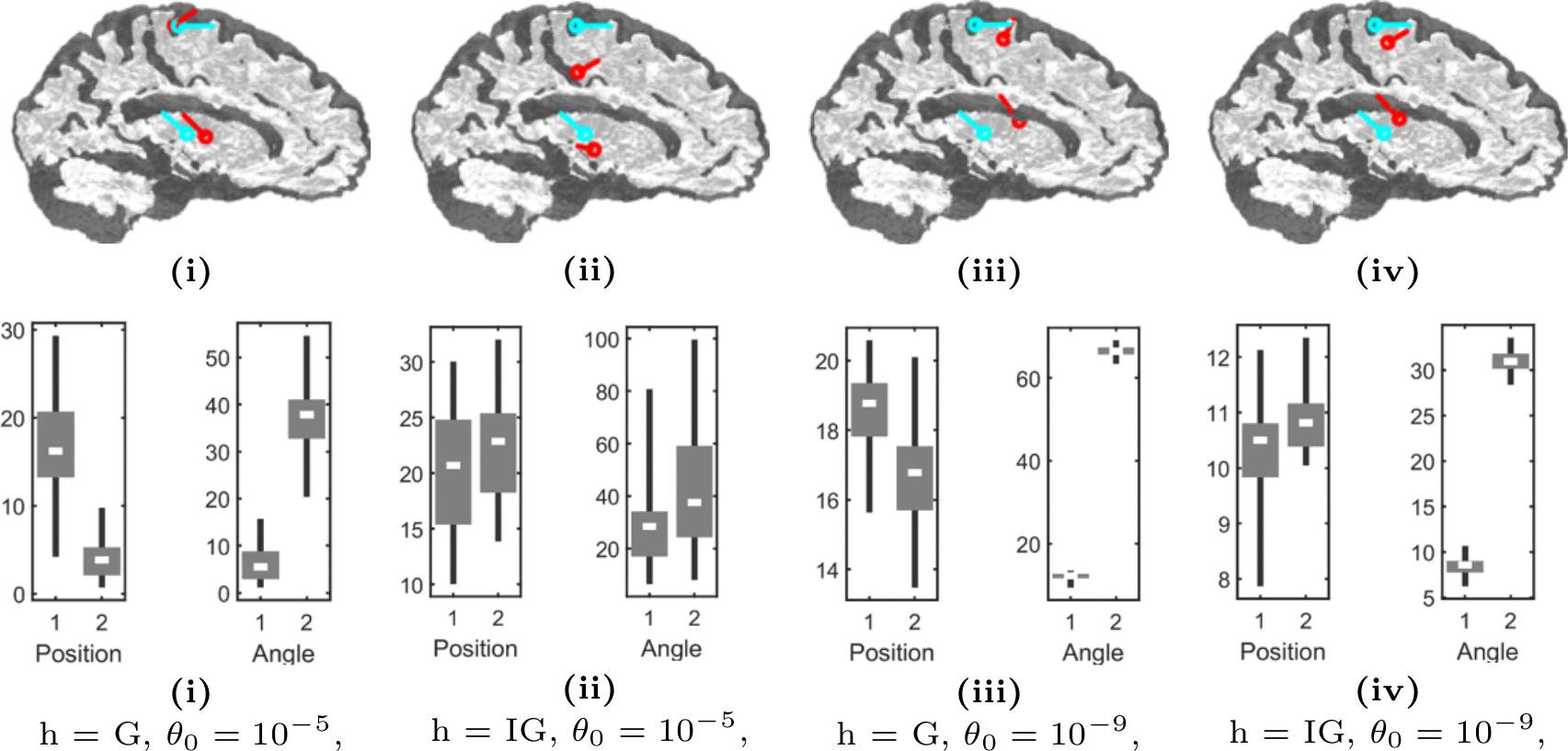}
  \caption{{\bf Top row:} Examples of the center of mass (red pin) found for the deep and superficial source (1 and 2, respectively) in the cases {(\bf i)}--{(\bf iv)} with synthetic EEG data and the noise level of 2 \%.  The exact position of each source is also depicted (cyan pin). {\bf Bottom row:} Box-plots showing the distributions {(\bf i)}--{(\bf iv)} of the position (millimeter) and angle error (degree) found for sources 1 and 2 and 50 different realizations of the noise vector.   }
  \label{roi_reconstructions}
\end{center}
\end{figure}

\section{Results}

ZI's forward simulation performance was evaluated w.r.t.\ the computing time for the head model described in Section \ref{hypermodel_selection}. The mesh generation, LF matrix evaluation and interpolation processes took 21, 39 and 3.5 minutes, respectively, using  NVIDIA\footnote{\url{https://en.wikipedia.org/wiki/List_of_Nvidia_graphics_processing_units}}  Quadro P6000 GPU. GPU acceleration was also found to be necessary to obtain a reasonable computing time as it sped up these routines by more than a factor of ten. 

\subsection{EEG inversion test} 

The results of the EEG inversion test can be found in Figure \ref{eeg_reconstructions} which displays  the reconstructed brain activity for the surface of the cortex and a volume cut corresponding to the  location of the maximal activity. The reconstruction obtained with the IG hyperprior was observed to be more focal than the one corresponding to G. The surface visualizations obtained for the G and IG hyperprior confirm similar active area as illustrated in Brainstorm's {\em EEG and epilepsy} tutorial, especially, compared to the outcome of the Brainstorm's maximum of entropy on the mean (MEM) framework solution. The volume cuts show the depth of  the reconstructed activity. 

\subsection{EIT inversion test} 

In the EIT inversion test, the averaged reconstruction found for the synthetic hemorrhage matched well with its exact location, which is shown in Figure \ref{eit_reconstructions}. A visual comparison between the averaged and unaveraged and reconstruction (Figure \ref{eit_reconstructions}) suggests that the averaging process was beneficial w.r.t.\ the localization accuracy. The resolution (level of detail) of the averaged reconstruction seems to be refined compared to that of the unaveraged one. Moreover, using a comparatively low number of DOFs (here 300) in each  randomized decomposition was found to be necessary for detecting  the hemorrhage.   

\subsection{Hypermodel and parameter selection}



Figure \ref{roi_reconstructions} illustrates the source localization results obtained in the hypermodel and parameter selection test.  G was observed to perform comparably well for the superficial ROI and IG for the deep one. This is reflected by the cases {(\bf i)} and {(\bf iv)}  in which the utmost position accuracy was obtained in these ROIs, respectively. Moreover, for G, the larger scale parameter value seemed preferable to the smaller one, while, for IG, the situation was the opposite. Regardless of the hyperprior, selecting a smaller scale parameter seemed beneficial for localizing the deep source  until a certain level, where noise effects started to affect the reconstruction.

\section{Discussion}

This article introduced {\em Zeffiro} interface (ZI) version 2.2, a GPU accelerated Matlab tool for multi-modal FEM-based modeling  of electromagnetic fields in brain imaging and investigations \citep{braess2001,demunck1988,monk2003}. It was shown that, when aided by a state-of-the-art GPU, ZI allows one to invert a given set of EEG data for a physiologically accurate \citep{rullmann2009eeg} one-millimeter volumetric multi-compartment head model within a reasonable one hour's time. GPU acceleration is needed, specifically, in the forward simulation phase, that is, in the generation of the FE mesh and the LF matrix as well as in the interpolation process connecting the DOFs of the unknown with the nodes of the FE mesh. Since Matlab does not currently parallelize the sparse matrix operations in a CPU, the performance difference between CPU and GPU computations, both applicable in ZI, is particularly pronounced.

As the mutual performance of GPU- and CPU-based codes is strongly system-specific and depends on various factors in addition to the processors themselves,  ZI was not directly compared to the alternative tools. These  include, for instance,  Duneuro\footnote{\url{http://duneuro.org}} \citep{nussing2019duneuro} and SimBio\footnote{\url{http://simbio.de}} \citep{fingberg2003bio} which are  open source FEM libraries for EEG/MEG with similar functions as Zeffiro but utilizing C++ language. 
Brainstorm\footnote{\url{https://neuroimage.usc.edu/brainstorm/Introduction}}  \citep{tadel2011brainstorm}  and Fieldtrip\footnote{\url{http://www.fieldtriptoolbox.org}} \citep{oostenveld2011} are alternative packages for the Matlab platform. The core forward modeling approach of Brainstorm is the BEM \citep{he1987electric}. Fieldtrip does not have an advanced forward and inverse modeling functions. None of these are currently capable of advanced  FEM or GPU computations. The MNE-Python\footnote{\url{https://martinos.org/mne/stable/index.html}} toolbox \citep{gramfort2013meg} is the leading option for Python. It allows utilizing a GPU, but is, nevertheless, limited to a BEM-based forward simulation. 

The present results suggest that ZI enables robust inversion of multi-modal data. 
Firstly, the reconstructions obtained for Brainstrom's {\em EEG and epilepsy} tutorial dataset show that ZI's forward and inversion methods can be applied to detect brain activity. Secondly, based on the numerical results obtained in the EIT inversion test, it seems that ZI can also be extended for non-linear problems and inversion of scalar-valued fields. The IAS MAP estimation technique was found to be applicable for EIT via averaging MAP estimates obtained for a randomized set of low-resolution domain decompositions. This technique might be usable also for other imaging modalities, for example, to localize deep brain activity. Furthermore, the present EIT solver might be adapted for other applications involving current injections, such as transcranial electric stimulation  \citep{herrmann2013transcranial} in which the brain activity is evoked through external stimuli. 


Based on the hyperprior and scale parameter selection experiment, IG seems to be an advantageous choice for the sub-cortical areas, whereas G seems preferable for the cerebral cortex. The scale parameter applied in the former case should be generally lower than in the latter one. This baseline is in parallel with the previous findings \citep{calvetti2009} and might be optimized later on. We also emphasize that the parameter selection is generally a complicated issue which is not covered completely in this study. For example, the effect of the shape parameter, which partially overlaps with that of the scale parameter, is omitted here. 

Compared to the BEM, the FEM has at least two major advantages when applied in EEG/MEG. Firstly, while a BEM solver slows down if the surface mesh resolution or the number of surfaces grows, the computational performance of the FEM is virtually independent of these factors. ZI's current design takes this aspect into account, as altogether 27 tissue compartments, each one composed of sub-entities if needed, can be included in a single head model. The uniform mesh generator is well-suited for multi-compartment meshing, since unlike many widely used software, e.g., TetGen\footnote{\url{http://wias-berlin.de/software/tetgen/}} \citep{tetgen} and Netgen\footnote{\url{https://ngsolve.org/}}  \citep{netgen}, it allows the tissue boundaries to intersect each other without collapsing. 
This is essential in practice, since the segmentation routines utilizing MRI data do not  always render the surfaces smoothly. Moreover, ZI performs appropriately with a high surface resolution, thereby, allowing one to  directly use the detailed surfaces exported from the  FreeSurfer\footnote{\url{https://surfer.nmr.mgh.harvard.ed}} Software Suite \citep{fischl2012freesurfer}.  

The second major benefit of the FEM is that the  conductivity distribution can be anisotropic  \citep{gullmar2010,rullmann2009eeg}. ZI's forward simulation routines are currently capable of handling  anisotropicity. The current meshing routine, however, generates an isotropic conductivity distribution. Generating or importing an anisotropic distribution, e.g., from diffusion-weighted MRI data, constitutes a potential topic for the future work. 

Another potential direction is to develop the inversion methodology: one might apply the HBM for sub-cortical areas \citep{seeber2019subcortical}, with non-diagonal prior covariance structures and/or with sampling-based posterior exploration techniques, e.g., the Gibbs sampler  \citep{spitzer1971markov,murphy2012machine}. From the practical viewpoint, there is also an obvious need to develop tools for various purposes including epochs, the covariance of the measurements, and connectivity, e.g., phase-locking \citep{lachaux1999measuring}. Clinical studies would be needed to validate ZI for different applications and measurement situations. Providing command line executable scripts for performing the main operations without the graphical user interface, e.g., in a computing cluster, is also a potential option.

\section*{Information Sharing Statement}

The results of the paper have been produced using the openly available  Zeffiro interface (ZI) code package$\mbox{}^{1}$ which has been implemented for the MATLAB (The MathWorks Inc., \url{https://www.mathworks.com}) environment. EEG source localization accuracy has been analyzed with the epilepsy tutorial dataset$\mbox{}^{4}$ of the Brainstorm$\mbox{}^{9}$ software with the consent of Prof.\ A.\ Schulze Bonhage, Epilepsy Center, University Hospital Freiburg, Germany. The EIT results have been computed with the population head model$\mbox{}^{5}$. The parcellation of the brain (36 Desikan-Killiany labels) has been obtained with the FreeSurfer software suite$\mbox{}^{2}$ and visualized with ZI.

\section*{Acknowledgments}

QH, AR, and SP contributed equally to this study. This study was supported by the Academy of Finland Centre of Excellence in Inverse Modelling and Imaging 2018--2025. We are grateful for DAAD's (Deutscher Akademischer Austausch Dienst) bilateral travel funding support between Mathematics and Statistics, Tampere University and Institute for Biomagnetism and Biosignal Analysis, University of Münster (Project: Advancing Finite Element Computations for Reconstructing and Manipulating the Human Somatosensory Cortex, Academy of Finland decision number 317165).

\appendix

\section{Finite elements in multimodal lead field evaluation}

To model electromagnetic fields, ZI applies the finite element method which allows obtaining lead field matrices for multiple different applications and data modalities. This appendix shows mathematically, how the lead field matrices of the EEG and linearized EIT problem are obtained in ZI, when the complete electrode model (CEM) is applied. 

\subsection{Complete electrode model in lead field evaluation}
\label{appendix_cem}

The governing PDE can be equipped with the following (lumped) CEM boundary conditions \citep{cheng1989}.  {\bf (I)}: $\sigma \nabla u \cdot \vec{n}|_{\partial \Omega \setminus \cup_\ell e_\ell}   = 
0$, {\bf (II)}: $\int_{e_\ell} \sigma \nabla u \cdot \vec{n} \, d S  =  I_\ell$,  and {\bf (III)}:  $( u + Z_\ell \mathcal{A}_{\ell} \sigma \nabla u \cdot \vec{n} )
|_{e_\ell}  =  U_\ell$
 for $\ell = 1, 2, \ldots, L$, where $\vec{n}$ denotes the surface normal. According to the first condition  {\bf (I)},  the normal current $\sigma \nabla u
\cdot \vec{n}$ on $\partial \Omega$ can flow out of or into the domain only through electrodes $e_\ell$, $\ell = 1, 2, \ldots, L$. The second one {\bf (II)} sets the net current flowing through  each electrode is $I_\ell$,  and the third one {\bf (III)} corresponds to the potential jump on the skin-electrode contact boundary. The voltage of the $\ell$-th electrode is denoted by $U_\ell$. ${Z}_\ell$ is the average contact impedance or resistance and $\mathcal{A}_\ell$ is the contact area of the $\ell$-th electrode. An additional condition is the the equation $\sum_{\ell = 1}^L I_\ell = 0$ which guarantees that the subject is grounded appropriately, so that there is no current flowing out of the head through the  neck. Integrating the governing PDE for the potential field, i.e., $\nabla \cdot (\sigma \nabla u) = \nabla \cdot \vec{J}^{\, p}$, by parts yields the for weak form \citep{pursiainen2016b}: 
{\setlength\arraycolsep{2 pt} \begin{eqnarray}
\label{weak_form}    - \!\! \int_\Omega  ( \nabla \cdot \vec{J}^{\, p} ) v \, d V & = & \int_\Omega  \sigma \nabla u \cdot
\nabla v \, d V   +     \sum_{\ell = 1}^L  \frac{1}{{Z}_\ell \mathcal{A}_\ell}\int_{e_\ell}  u \, v \, d S
  \nonumber\\  &  & -   \sum_{\ell = 1}^L
\frac{1}{{Z}_\ell \mathcal{A}_\ell^2 } \int_{e_\ell}  u \, d S \int_{e_\ell}  v \, d
S  - \sum_{\ell=1}^L Z_\ell I_\ell.
\end{eqnarray}}
If the divergence of $\vec{J}^{\, p}$ is square integrable, i.e., if $\vec{J}^{\, p} \in \{ \vec{w} \, | \,   \nabla \cdot \vec{w} \in L^2(\Omega) \}$, the weak form  has a unique solution $u \in H^1(\Omega)=\{ w \in L^2(\Omega)$ : ${\partial w}/{\partial x_i } \in L^2(\Omega)$,  $i = 1,2,3 \}$ satisfying (\ref{weak_form}) for all $v \in H^1(\Omega)$.  The weak form  (\ref{weak_form}) can be discretized in a straightforward way via the classical Ritz-Galerkin technique \citep{braess2001} which yields the system
 \begin{equation}
\label{u_system} \left( \begin{array}{cc} {\bf A} & -{\bf B} \\
-{\bf B}^T & {\bf C}
\end{array} \right) \left( \begin{array}{c} {\bf z}  \\
{{\bf v}}
\end{array} \right) = \left( \begin{array}{c} - {\bf G} {\bf x}  \\
{\bf I}
\end{array} \right).
\end{equation}
Matrix ${\bf A}$ is of the form
$
{a}_{i, j}  \! = \! \int_{\Omega} \! \sigma \nabla \psi_i \cdot \nabla \psi_j \, d V \! + \!
\sum_{\ell = 1}^L \frac{1}{Z_\ell \mathcal{A}_\ell} \! \int_{e_\ell} \psi_i \psi_j \, dS,
\label{tupu}
$
where $\psi_i$, $i = 1,2,\ldots,n$ are linear (nodal) FE basis functions. To ensure the invertibility of  ${\bf A}$, it is additionally defined that the identities ${a}_{i',i'} = 1$ and ${a}_{i',j} = 0$ ($j \neq i'$) are satisfied for the index $i'$ corresponding to  a  basis function $\psi_i'$  which is maximized on the boundary $\partial \Omega \setminus \cup_\ell e_\ell$ not covered by the electrodes. The entries of ${\bf B}$, ${\bf C}$ and ${\bf G}$ are given by $b_{i, \ell}   =  \frac{1}{Z_\ell} \int_{e_{\ell}} \psi_i \, dS$, $c_{\ell, \ell}   =  \frac{1}{Z_\ell} \int_{e_\ell} \,  dS$, 
${c}_{i, \ell} = 0$ ($i \neq \ell$), and  $g_{i,j}  = \int_\Omega  \psi_i (\nabla \cdot {\vec w}_j )  d V$, where ${\vec w}_j$, $j = 1,2,\ldots,m$ are basis functions belonging to the $H(\hbox{div})$ space.  The current vector ${\bf I} = (I_1, I_2, \ldots, I_L )$ is nonzero, if the electrodes are actively injecting currents. The zero-mean electrode  voltage vector ${\bf y} = (U_1, U_2, \ldots, U_L)$ predicted by (\ref{u_system}) can be obtained via ${\bf y} = {\bf
R} {\bf v}$ in which the matrix ${\bf R}$ defined by $r_{j,j} = 1 - 1/L$ for $j = 1, 2, \ldots,L$, and
$r_{i,j} = -1/L$  ($i \neq j$) . 

\subsection{EEG lead field}

In EEG, the electrode currents included in ${\bf I}$ are  zero, as the electrodes only measure the voltage on the skin. Thus, vector ${\bf v}$ can be explicitly solved from (\ref{u_system}) which leads to the expression
$ {\bf y} = {\bf R} ( {\bf B}^T {\bf A}^{-1} {\bf B} -
{\bf C} )^{-1} {\bf B }^T {\bf A}^{-1} {\bf G} {\bf x}$, 
and, further, to the following EEG LF matrix:
\begin{equation}
\label{aku} {\bf L}  =   {\bf R} ( {\bf B}^T
{\bf A}^{-1} {\bf B} - {\bf C} )^{-1} {\bf B }^T {\bf A}^{-1} {\bf
G}.
\end{equation}
The lead field of the MEG problem can be derived in an analogous way using the Biot-Savart formula for the magnetic field as shown in \citep{pursiainen2012raviart}.

\subsection{Linearized EIT lead field}

In EIT, the primary current density can be assumed to be zero, as the magnitude of the injected currents is far  superior to the brain activity. The unknown of the EIT inverse problem is the conductivity distribution $\sigma$. The voltage measurements ${\bf y} = {\bf R} {\bf v}$ generated by the current injections ${\bf I}$ are used as the data. The forward model that follows is given by 
$ 
    {\bf y} = {\bf R} {\bf M}^{-1} {\bf I}, 
$ 
where ${\bf M} = ({\bf C} - {\bf B}^T  {\bf A}^{-1} {\bf B})$. 
The conductivity distribution is assumed to be piecewise (element-wise) constant, i.e., of the form $\sigma =   \sum_{m=1}^{M} s_m \chi_m$, where  $\chi_m$ is the indicator function of  the element $m$ in the FE mesh. Denoting by $\sigma^{(bg)}$ a background conductivity distribution, i.e., the point of linearization, the unknown of the inverse problem is the difference vector ${\bf x} = (s_1-s_1^{(bg)}, s_2-s_2^{(bg)}, \ldots, s_M-s_M^{(bg)})$.   The LF for linearized EIT can be derived by differentiating both sides of the equation ${\bf M} {\bf v} = {\bf I}$ as follows:
$
{\bf 0} = -  {\bf B}^T \frac{\partial}{\partial s_m} {\bf A}^{-1} {\bf B} {\bf v} + {\bf M} \frac{\partial}{\partial s_m} {\bf v}.  
$ 
 Moreover, a straightforward differentiation of the equation ${\bf A} {\bf A}^{-1} = {\bf I}$ shows that  $ \frac{\partial}{\partial s_m} ( {\bf A} {\bf A}^{-1}) =    \frac{\partial {\bf A} }{\partial s_m}   {\bf A}^{-1} +   {\bf A} \frac{\partial {\bf A}^{-1}}{\partial s_m}  = {\bf 0}, 
$
and, further, that $\frac{\partial {\bf A}^{-1}}{\partial s_m} = - {\bf A}^{-1} \frac{\partial {\bf A} }{\partial s_m}   {\bf A}^{-1}$.
Taking into account that $\frac{\partial {\bf y}}{\partial s_m}  = {\bf R} \frac{\partial {\bf v}}{\partial s_m} $, the linearized lead field can be written as
$
 \frac{\partial {\bf y}}{\partial s_m}  = - {\bf R}  {\bf M}^{-1}  {\bf B}^T {\bf A}^{-1} \frac{\partial {\bf A} }{\partial s_m}   {\bf A}^{-1} {\bf B} {\bf v}. 
$
Thus, the differential is of the form 
\begin{equation}
 \frac{\partial {\bf y}}{\partial s_m}  = - {\bf R} {\bf M}^{-1}  {\bf B}^T {\bf A}^{-1} \frac{\partial {\bf A} }{\partial s_m} {\bf A}^{-1} {\bf B} {\bf M}^{-1} {\bf I}.
\end{equation} 
The linearized forward model of EIT is given by ${\bf y} \approx {\bf L} {\bf x}  + {\bf y}^{(bg)}$, where ${\bf y}^{(bg)}$ is a simulated data vector corresponding to the background conductivity distribution ${\sigma}^{(bg)}$ and the  entries of the lead field matrix ${\bf L}$ are of the form ${l}_{k,m} = \partial y_k /\partial s_m |_{{\sigma}^{(bg)}}$. 

\subsection{Transfer matrix}
\label{transfer_matrix}

Both EEG and EIT lead field matrix can be formed by first evaluating the so-called transfer matrix ${\bf T} = {\bf A}^{-1} {\bf B}$. Obtaining a single column ${\bf t}$ of ${\bf T}$ necessitates solving a linear system of the form ${\bf A} {\bf t} = {\bf b}$, where ${\bf b}$ is a single column of the  matrix ${\bf B}$ which has as many columns as there are electrodes in the measurement system.

\section{IAS MAP inversion}
\label{appendix_ias}

The iterative alternating sequential (IAS) inversion approach \citep{calvetti2007gaussian,calvetti2009,calvetti2018bayes} to find a {\em maximum a posteriori} estimate for the posterior density is given by:
\begin{enumerate}
\item Choose parameters $\beta$ and $\theta_0$. Set $k = 1$ and ${\bm \theta}^{(0)} = (\theta_0, \theta_0, \ldots, \theta_0)$.
\item Find ${\bf x}^{(k)} = \arg \max_{x} p ({\bf x}  \mid {\bf y}, {\bm \theta^{(k-1)}})$.
\item Find ${\bm \theta}^{(k)} = \arg \max_{\bm \theta} p ({\bm \theta} \mid {\bf y}, {\bf x}^{(k)})$.
\item If $k$ is less than the total number of iterations chosen by the user, then go to 2.\ and set $k = k +1$, else set ${\bf x}_{MAP} = {\bf x}^{(k)}$.
\end{enumerate}
IAS finds a conditional maximum of the posterior alternatingly with respect to the unknown vector ${\bf x}$ and the hyperparameter ${\bm \theta}$.   The algorithm can be, further, written as
 \begin{enumerate}
\item Set $k=0$ and ${\bm \theta}^{(0)} = (\theta_0, \theta_0, \ldots, \theta_0)$.
\item Set ${\bf L}^{(k)} = {\bf L} {\bf D}^{1/2}_{{\bm \theta}^{(k)}}$ with $ {\bf D}^{1/2}_{{\bm \theta}^{(k)}}= \hbox{diag} (\sqrt{ |{\bm \theta}^{(k)}_1|}, \sqrt{|{\bm \theta}^{(k)}_2|}, \ldots, \sqrt{|{\bm \theta}^{(k)}_n|}). $
\item Evaluate \begin{equation}
{\bf x}^{(k+1)} =  {\bf D}^{1/2}_{{\bm \theta}^{(k)}} {{\bf L}^{(k)}}^T ( {\bf L}^{(k)} {{\bf L}^{(k)}}^T + \nu^2  {\bf I})^{-1} y,  \label{ias_matrix} \end{equation}
 where $\nu$ denotes the standard deviation of the likelihood.
\item Update the hyperparameter based on the hypermodel. 
\begin{itemize} 
\item If the hypermodel is G, set $ \theta_i = \frac{ 1 } { 2 }  \theta_0 \Big( \eta + \sqrt{\eta^2 + 2 {x_i^{(k)}}^2/\theta_0} \Big) $ with $\eta = \beta - 3/2$, $i = 1, 2, \ldots, n$.
\item Else, if the hypermodel is IG, set $ \theta_i^{(k+1)} =  (\theta_0 +  \frac{{x_i^{(k)}}^2}{2})/\kappa $ with $\kappa = \beta + 3/2$, $i = 1, 2, \ldots, n$.
\end{itemize}
\item Set $k = k +1$ and go back to 2., if $k$ is less than the total number of iterations defined by the user. \end{enumerate}


\end{document}